\documentclass{article}

\usepackage{latexsym}
\usepackage{amsmath}
\usepackage{amssymb}
\usepackage{url}

\newcommand{\at}{\mathtt{@}}

\newcommand{\ife}{\mathit{ife_{ft}}}
\newcommand{\life}{\mathit{life_{ft}}}

\newcommand{\ifm}{\mathit{ifm_{ft}}}

\newcommand{\fifg}{\mathit{fifg_{ft}}}
\newcommand{\lifm}{\mathit{lifm_{ft}}}
\newcommand{\ifgB}{\mathit{ifg_{ft,B}}}
\newcommand{\fifgB}{\mathit{fifg_{ft,B}}}

\newcommand{\ifmB}{\mathit{ifm_{ft,B}}}

\newcommand{\lifmB}{\mathit{lifm_{ft,B}}}

\newcommand{\less}{\leq}

 \newdimen\boxwdplusemdimen
  \def\arrow#1{{
    \boxwdplusemdimen=1em%
    \setbox0=\hbox{$\scriptstyle#1$}%
    \advance\boxwdplusemdimen by \wd0\relax%
    \ifdim\boxwdplusemdimen<16.11119pt%
      \boxwdplusemdimen=16.11119pt%
    \fi%
    \buildrel{#1}\over%
      {\setbox1=\hbox to \boxwdplusemdimen{\rightarrowfill}%
    \ht1=0.35em\relax\box1}
  }}

\begin{document}

\title{Interface groups and financial transfer architectures
\thanks{An earlier version
of this paper appeared as report PRG0702, Section
Software Engineering, Informatics Institute,
Faculty of Science, University of Amsterdam.}}
\author{Jan A. Bergstra$^{1,2,}$\thanks{Jan Bergstra acknowledges support
from NWO Jacquard project Symbiosis.} \and Alban Ponse$^1$\\[1ex]
\small
\begin{tabular}{l}
${}^1$University of Amsterdam, Section Software Engineering,\\
Kruislaan 403, 1098~SJ~Amsterdam, The Netherlands\\[1ex]
${}^2$Utrecht University, Department of Philosophy,\\
Heidelberglaan 8, 3584~CS~Utrecht, The Netherlands
\end{tabular}
}

\date{11 July 2007}
\maketitle

\begin{abstract}
Analytic execution architectures have been proposed by the same authors
as a means to conceptualize the cooperation between heterogeneous
collectives of components such as programs, threads, states and
services. Interface groups have been proposed as a means to formalize
interface information concerning analytic execution architectures.
These concepts are adapted to organization architectures with a focus
on financial transfers.  Interface groups (and monoids) now provide a
technique to combine interface elements into interfaces with the
flexibility to distinguish between directions of flow dependent on
entity naming.

The main principle exploiting interface groups is that when composing a
closed system of a collection of interacting components, the sum of
their interfaces must vanish in the interface group modulo reflection.
This certainly matters for financial transfer interfaces.

As an example of this, we specify an interface group and within it some
specific interfaces concerning the financial transfer architecture for
a part of our local academic organization.

Financial transfer interface groups arise as a special case of more
general service architecture interfaces.

\end{abstract}

\section{Introduction}

In \cite{BP06} we proposed ``analytic execution architectures'' as a
means to conceptualize the cooperation between key components such as
programs, threads, states and services. Interfaces are a practical tool
for the development of all but the most elementary architectural
designs. We will now use that terminology as well for the case that
components mainly interact by transferring financial assets amongst
one-another.

Interface groups have been proposed in \cite{BP06b} as a technique to
combine interface elements into interfaces with the flexibility to
distinguish between permission and obligation and between promise and
expectation which all come into play when component interfaces are
specified.

As a vehicle to present and investigate interface groups
we have used the program algebra PGA as defined in
\cite{BL02} and thread algebra (TA,~\cite{BM04},~\cite{PZ}).

From the set of basic actions $A$ that underlies any program algebra or
thread algebra a set $\ife(A)$ of interface elements is derived.  These
generate the interface group for $A$ (in additive notation).  In the
case of a financial transfer architecture the rather simplistic
assumption is that an organization is composed of a number of
entities.  Assuming that this decomposition into entities is stable for
a significantly extended period of time, it becomes  both meaningful
and helpful to specify for each entity in an organization to what
extent it  can make financial transfers to other entities inside or
outside the organization. Below interface groups are proposed as a
means to specify large sets of interfaces from which appropriate ones
may be chosen to represent a certain observed or imagined 
financial transfer architecture.

The main principle that makes use of an interface group is that when
composing a closed system of a collection of interacting components,
the sum of their interfaces must be $0$. This holds in the case of an
interface architecture for a program execution architecture just as
much as for a financial transfer architecture that is supposed to shed
light on some complex financial transfer setting.



What has been left out on purpose
 at this level of abstraction is what many people seem to consider the
most important:  quantitative information. The idea is that each pair
of entities and direction and even each 'explanation' or motive may
entail different rules of engagement which may be needed to decide or
compute how a certain transfer is to be achieved. In some cases
additional or complementary payments may be needed to other parties
which one may not yet know or which one prefers to hide at a certain
level of abstraction. For instance taking money from a cash point might
involve a transfer towards one's bank which is hidden at some level of
abstraction. In general, transaction costs may be preferably ignored at
initial stages of the design of a financial transfer architecture only
to be specified in a subsequent stage of refinement. Similarly designs
may initially ignore theft, fraud or misuse of services, only to add
these 'features' in subsequent refinement stages.

The paper is structured as follows:
in Section~\ref{sec:2} we introduce interface elements and
various kinds of interface groups.
Then, in Section~\ref{sec:3} we consider localization,
globalization, and some other natural operations on
interface groups.
In Section~\ref{sec:4} we introduce components and
architectures for describing financial transfers.
Then, in Section~\ref{sec:5} we discuss the design of some financial
transfer interfaces by examples.
In Section~\ref{sec:6} we provide some discussion and concluding
remarks.

\section{Interface elements, interface monoids and interface groups}
\label{sec:2}
In this section we introduce our basic technical ingredients:
interface elements and interface groups. 
When working on the design of a financial transfer architecture,
it is suggested that one is very precise about the interfaces of
the components
and that all interfaces are chosen as elements of an interface
group. We shall often write ``FT-interface'' instead of ``financial
transfer interface".

\subsection{Financial transfer interface elements}
Three ingredients are presupposed in the definition of 
FT-interface elements:
\begin{enumerate}
\item
A finite set $A$ of so-called basic transfer actions, these will carry 
information about the form of transfer, e.g., cash, electronic wallet, 
credit card, debit card, bank transfer, annual, monthly, weekly, daily,
random.
\item
A finite set $E$ of entities which will
serve as components of financial transfer architectures.
In our example below these will be various parts
of a Faculty of Science of a Dutch University.
\item
A finite set $M$ of motives or explanations. A motive explains why a 
transfer is made, the typical examples being `salary' or
`travel reimbursement'.
\end{enumerate}

\noindent
The set of \emph{FT-interface elements} is introduced: 

\begin{align*}
\ife(E,A,M)) &= \{e.a(m)\at f/\alpha\mid e, f \in E,~a \in A,~m \in
M,~\alpha \in  \{TF,T,F,\lambda\}\}\\
&\cup \{{\sim} e.a(m)\at f/\alpha\mid
e, f \in E,~a \in A, m \in M,~\alpha \in  \{TF,T,F,\lambda\}\}.
\end{align*}

\noindent
The intended meaning of these interface elements is as follows:

\begin{itemize}
\item
$e.a(m)\at f/TF$ indicates the permission (option, ability)
of an entity $f$ to issue
a financial transfer  $a(m)$ (action $a$ with motive $m$)
towards an entity $e$, while 
expecting reply either $T$ or $F$ with $T$ representing that
the request has succeeded
and the transfer has taken place and $F$ representing a
refusal of the transfer by $e$. 
The element
$e.a(m)\at f/TF$ is called a \emph{service interface element}
or \emph{outgoing transfer element}.

\item
${\sim} e.a(m)\at f/TF$ indicates the permission (option, ability)
of a component $f$ 
to receive a transfer $a(m)$ from entity $e$, where $f$ has the
right to either accept
or refuse which is is signaled by returning reply $T$ or $F$ to
the issuing entity $e$.
The element
${\sim} e.a(m)\at f/TF$ is called a \emph{client interface element}
or \emph{incoming transfer element}.
\end{itemize}

If $\alpha$ is $T$ the request must be accepted, if  $\alpha$ is $F$
the request is always refused and if  $\alpha$ is $\lambda$ no
information is given about whether or not the request is accepted.
Here $\lambda$ denotes the empty string.

In the remainder of this paper all interface elements used will be of the
form $e.a(m)\at f/TF$ or ${\sim} e.a(m)\at f/TF$.
This implies that
components are never forced to accept incoming transfers or to permit
outgoing transfers. Such decisions are made dynamically. For instance
if an interface describes transfers during some standardized time slot
(e.g. 24 hours) it is possible that a certain transfer is accepted only
once, while all subsequent requests are turned down.
By restricting $\alpha$ to $TF$ the possibility to indicate that some
transfers are always accepted and others may be never accepted is given
up.  The features can be used to express that some components are in
the lead, because some other components will always accept their
transfer request, or to express that a very plausible request will
never be accepted. The difference between an interface containing
$e.a(m)\at f/F$ and the same interface not containing $e.a(m)\at f/F$
is that the second interface considers a transfer $e.a(m)\at f$ a
static error whereas in the first case it is considered a dynamic
failure.
We will use the following abbreviation:
\[e.a(m)\at f = e.a(m)\at f/TF.\]

In this paper
use will be made of interface combination, notation $+$, and interface
inversion, notation $-$.
The intended meaning of interfaces derives
from the intended meaning of client and service interface elements as
described above.
In a composed interface $I+J$ it is implied that the combination of two
options is the option (ability, permission) to do both. 
To 
simplify the notation the following ordering of precedence is used: 
\[+ < - < / < \at < {\sim} < .
\]
 Moreover the convention $X-Y = X + (-Y)$ is used. 
For instance, 
\[I - {\sim} f.a(m)\at g -J\quad\text{stands for}\quad
I + (-(({\sim}(f.a(m)))\at g)) + (-J).\]
Instead of $a(m)$ we will write often  $a_m$ in order to save brackets.

It is purely a matter of design to take $f.a_m\at g$ to represent a
transfer to $f$ rather than from $f$. In fact two decisions are
implicit in the notation:
\begin{enumerate}
\item $f.a_m\at g$ rather than
$g.a_m\at f$ represents a transfer to $f$ made by $g$ because
$g$ performs the action which changes the state of $f$ (provided $f$
accepts the transfer),
\item $f.a_m\at g$ rather than
$-f.a_m\at g$ represents a transfer made by $g$ because it will
usually be on the initiative of $g$ that its transfer to $f$ is made.
That initiative, however, may well take place as a consequence of some
preceding request for this transfer expressed in a different notation.
\end{enumerate}

\subsection{The financial transfer interface monoid}
\label{ftig}
The \emph{FT-interface monoid} $\ifm(E,A,M)$ is the
commutative monoid with additive notation, generated from the set
$\ife(E,A,M)$.

With $\ifmB(E,A,M)$ the submonoid of
$\ifm(E,A,M)$ is denoted which is generated by interface elements of
the form  $e.a(m)\at f/TF$ and ${\sim} e.a(m)\at f/TF$ (here $B$
stands for the set $\{T,F\}$). Working in
$\ifmB(E,A,M)$ we stick to the abbreviation $e.a_m\at f$ for
$e.a(m)\at f/TF$.

\subsection{Financial transfer interface groups}
\label{ftim}
Three groups will be used for calculation with FT-interfaces. The free
FT-interface group permits incremental modifications of interfaces.

\subsubsection{The free financial transfer interface group}
Interface architecture descriptions can be presented as elements of
$\ifmB(E,A,M)$. When modifying such descriptions is it useful to be
able to add and subtract interface elements. For that reason the
FT-interface monoid $\ifmB(E,A,M)$ is embedded in a free FT-interface group
$\fifgB(E,A,M)$. This is the free commutative group in additive
notation generated from the same generators as $\ifmB(E,A,M)$.

One may wonder what meaning can be assigned to the multiple occurrence
of an interface element in the free FT-interface group. Consider $f.a_m
\at g\,+\,f.a_m \at g$.  This interface may arise as the result of the
application of an abstracting homomorphism $\psi$ to an interface
$f.a_{m1} \at g\,+\,f.a_{m2} \at g$, where $\psi$ forgets the
distinction between $m1$ and $m2$. Non-trivial interface element
multiplicities therefore indicate that certain transfers may be
performed in different ways, from which an abstraction is made.

\subsubsection{The financial transfer interface reflector group}
The FT-interface reflector group $R$ is a subgroup of the free FT-interface
group which contains all interface elements that vanish if the
following equation (reflection law) is assumed: 
\[f.a_m \at g + {\sim} g.a_m\at f = 0.\]
Reflector elements  are interfaces of the form $f.a_m\at g\, +
{\sim} g.a_m  \at f $ and the reflector group is the subgroup of the
free FT-interface group generated by all reflector elements.
Note that interface elements of the particular form
$f.a_m \at f$ and ${\sim} f.a_m\at f$ are also in $R$.

\subsubsection{The financial transfer interface group modulo reflection}
Because the reflector group is a normal subgroup of the free FT-interface
group one may introduce the quotient group of both. This group, given
by
\[\fifgB(E,A,M)/R\]
is called the \emph{FT-interface
group modulo reflection}.  Obviously $\fifg(E,A,M)/R$ is the commutative
group in additive notation, generated from the set $\ife(A)$ as
generators, modulo the reflection law. The homomorphism from
$\ifgB(E,A,M)$ to $\fifgB(E,A,M)/R$ is called the \emph{reflection
 mapping} and is denoted with $\phi_R$.

The reflection law
$f.a_m\at g + {\sim} g.a_m\at f = 0$ holds in the FT-interface group
modulo reflection. It can be written equivalently as 
\[-f.a_m\at g = {\sim} g.a_m\at f\quad\text{or}\quad  
f.a_m\at g = - {\sim} g.a_m\at f.\]

\subsection{Closed interfaces}
The main purpose of the introduction of $\fifgB(E,A,M)/R$ is that it
permits the following architectural integrity check on an interface
specified in $\fifgB(E,A,M)$:  $I \in \fifgB(E,A,M)$ is a closed
interface if $\phi_R(I) = 0$ in $\fifgB(E,A,M)/R$.

The motivation for having explicitly available
the monoid $\ifmB(E,A,M)$ and the free group
$\fifgB(E,A,M)$ rather than merely the quotient
$\fifgB(E,A,M)/R$ is twofold: the monoid contains interface
descriptions in normal form (with positive and negative elements
cancelled out). The free group is more expressive in permitting the
notion of changes (delta's) between different designs stages. Both the
monoid and the free group permit transitions back (localization) and
forth (globalization) to the localized interface monoids as introduced
in the following section. In particular, localization cannot be  defined
from the interface group modulo reflection. Because localization and
globalization are considered indispensable tools for understanding
complex interface descriptions this renders the use of the free group
unavoidable.

\subsection{Ordering and other structure}
Following \cite{BP06b} a partial ordering $\less$ on interfaces in
$\ifmB(E,A,M)$ is generated by these rules:

\begin{itemize}
\item
$0 \less p$ for all interface elements $p\in\ife(A)$,
\item
$0\less X $ if and only if $-X\less 0$,
\item
$X\less X+Y $ if and only if $0\less Y$.
\end{itemize}

Interfaces as modeled by interface groups have less structure than the
signatures used as interfaces in the module algebra of \cite{BHK90}. 
Module algebra, however fails to provide any concept of reflection
and for that reason it has a bias in the direction of the combination
of services (rather than clients). Module algebra and
similar approaches fail to provide the basic technical ingredients
needed for the description of analytic execution architectures which
are meant to combine various components such as clients and services in
asymmetric ways.

\section{Localization and globalization}
\label{sec:3}

For a particular entity it is unhelpful to always indicate
it being the source of 
outgoing transfers or the receiver of incoming transfers.
For that reason the
following notation is proposed:
\[\life(E,A,M) = \{f.a(m)/\alpha,  {\sim} f.a(m)/\alpha\mid
f \in E,~a \in A,~m \in M,~\alpha \in  \{TF,T,F,\lambda \}\}.
\]
These are called localized interface elements.
The intended meaning of these elements is as follows:

\begin{itemize}
\item
$f.a(m)/TF $ indicates the permission (option, ability) of
an entity (with its name left implicit, i.e.  a default entity) to issue
a financial transfer  $a(m)$
(action $a$ with motive $m$) towards an entity $f$, and
to expect either $T$ or $F$ as a reply respectively
signaling success of failure of the request.

\item
${\sim} f.a(m)/TF$ indicates the permission (option, ability)
of a default entity
to receive a transfer $a(m)$ from entity $f$ and
to reply either positively or negatively.
\end{itemize}

Like in the localized case with restricted forms of $\alpha$ the
corresponding restrictions on replies are assumed, and equally similar
to the non-localized case $f.a_m$ abbreviates $f.a(m)/TF$.

The \emph{localized financial transfer interface monoid} $\lifm(E,A,M)$
is the commutative monoid (in additive notation) generated by
$\life(E,A,M)$.

With $\lifmB(E,A,M)$ we denote the submonoid of  $\lifm(E,A,M)$
generated by interface elements of the form $f.a(m)/TF$ and ${\sim}
f.a(m)/TF$ which are in that context always abbreviated by $f.a_m$ and
${\sim} f.a_m$.

When working on the design of a financial transfer architecture it is
now suggested that for various entities
component interfaces are specified as elements of 
$\lifmB(E,A,M)$.

\subsection{From local to global and back}
Local  interfaces specify an interface from the perspective of a single
entity of which the name is left implicit whereas global interfaces
take a number of entities into account and contain all interface
information in a form which makes implicit names explicit.
Mathematically, local FT-interfaces exist in a free additive monoid,
and in a free additive group whereas global FT-interfaces exist in the
monoid as well as the free interface group and in addition to these in
the interface group modulo reflection from the previous section.

\subsection{Globalization}
For each entity $e \in E$ the mapping $\phi_e$ is a homomorphism from 
$\lifmB(E,A,M)$ into $\ifmB(E,A,M)$ given by the following equations for 
interface elements:
\[\phi_e(f.a_m)= f.a_m\at e\]
and
\[\phi_e({\sim} f.a_m)=  {\sim} f.a_m \at e.\]
The mapping $\phi_e$ is called \emph{globalization} as it turns a local
interface into a global one by making its implicit entity name
explicit.

\subsection{Localization}
In the opposite direction to globalization
\emph{localization} transforms global interfaces
(represented in $\ifmB(E,A,M)$)
to localized ones: $\overline{\phi_e}$.

Its defining equations are: 
\begin{eqnarray*}
\overline{\phi_e}(0)&=& 0,\\
\overline{\phi_e}(x + y)&=& \overline{\phi_e}(x) + \overline{\phi_e}(y),\\
\overline{\phi_e}(f.a_m\at g)&=& f.a_m \lhd (g=e)\rhd 0,\\
\overline{\phi_e}({\sim} f.a_m \at g)&=&{\sim}f.a_m \lhd (g=e)\rhd 0.
\end{eqnarray*}
Here $P \lhd c \rhd Q$
is the well-known infix alternative notation for the conditional
expression $\textit{if }c \textit{ then }P \textit{ else }Q$
which features in many program notations. It
is immediate that localization is a right inverse of globalization
on local interface elements:
$\phi_e \circ \overline{\phi_e} = Id_e$. 

Localization can be extended to all of $\fifgB(E,A,M)$ by means of 
these additional defining equations:
\begin{eqnarray*}
\overline{\phi_e}(- f.a_m\at g) &=& \overline{\phi_e}({\sim} g.a_m\at f),\\
\overline{\phi_e}(- {\sim} f.a_m \at g) &=& \overline{\phi_e}(g.a_m\at f).
\end{eqnarray*}

\subsection{Global interface decomposition}
\label{gigd}
Let $E$ be finite. For each interface $x \in \fifgB(E,A,M)$,
\[x = \sum_{e \in E}^{} \phi_e(\overline{\phi_e}(x)).\]
This representation
provides a systematic means to use local interface element notation
only and to present a large interface as a sum of globalized and
previously defined localized interfaces. These localized interfaces can
be designed at a stage where not even all relevant entities $E$ are
known.  This representation provides a decomposition of a global
interface into localized ones.

\subsection{Conditional interface elements}
Suppose that when designing a financial transfer 
interface it is unclear whether or not a
certain transfer may ever materialize. Then it can be helpful to use a
global boolean variable $c$ and a conditional local interface element 
\[ f.a(m) \lhd c\rhd 0\]
at entity $g$ while at
the complementary entity $f$ one  uses $-g.a(m) \lhd c\rhd 0$.
Whatever the boolean value of $c$ the 0-sum
condition need not be violated when working this way.

\subsection{Entity refinement homomorphisms}
It is reasonable to view entities as objects that can coexist in
parallel. Therefore the parallel composition $e\|f$ can be considered
an entity as well. Parallel composition is assumed to be associative
and commutative. Viewing an interface description within an interface
group as  a design stage it is reasonable to expand entity $f$ into
$f_1\|\ldots\|f_n$ thus expressing that $f$ consists of $n$ entities at
a lower level of abstraction. The homomorphism $\phi_{f\rightarrow
f_1\|\ldots\|f_n}$ works as follows on global interface elements (taking
$n=2$ for readability):\\
\[
\begin{array}{l}
\phi_{f\rightarrow f_1\|f_2}(g.a_m\at h) =\\ 
~~~~((f_1.a_m\at f_1 + f_1.a_m\at f_2 + f_2.a_m\at f_1 + f_2.a_m\at f_2)
 \\
 ~~~~\lhd (f = h) \rhd \\
 ~~~~(f_1.a_m\at h + f_2.a_m\at h) 
)\\
\lhd(f=g)\rhd\\ 
  ~~~~((g.a_m\at f_1+g.a_m\at f_2)
      \lhd (f=h) \rhd 
   g.a_m\at h),
\end{array}
\]
and in the same style an equation can be given for elements of
the form ${\sim}g.a_m\at h$.
After an application of a refinement homomorphism many options may
emerge that will play no further role. A further annihilation
homomorphism may then be needed to equate each irrelevant option with
the interface group unit $0$.

\subsection{Composition of motives}
To shorten the interface specifications it is helpful to
have a combination operator
$+$ on $M$ as well as on interfaces. The operator $+$ is assumed to be 
associative and commutative.
This turns the set of motives into a finitely 
generated free semi-group. 
 The following equations
axiomatize what is expected of mode composition in relation
to interfaces.
\begin{eqnarray*}
f.a(0) \at g &=& 0,\\
{\sim}f.a(0) \at g &=& 0,\\
f.a(v + w) \at g &=& f.a(v) \at g + f.a(w) \at g,\\
{\sim}f.a(v + w) \at g &=&{\sim}f.a(v) \at g +{\sim}f.a(w) \at g.
\end{eqnarray*}

\section{Combining components and describing architectures}
\label{sec:4}

Several terms are used to indicate the working of components in a
system.  In \cite{BP06b} we used interface elements with the additional
structure that subsequent to an action the service produces a boolean
reply value.
Architectural components that may implement such
interfaces are programs, program objects, instruction sequences and
polarized processes following the formalization of \cite{BL02,BB03},
and  threads, services and multi-threads as presented in \cite{BM04}.

What these terms have in common is that they make reference to
descriptions of the functionality (behavior, inner structure,
underlying mechanism) of parts of conceivable systems.  These parts are
either named by their role (thread, client or service) or by their
mathematical identity (process, program object, polarized process).

It is tempting to view these references as references to actual,
potential, designed or contemplated system components but we will
propose not to do so. Instead we will propose to view a component as a
pair $[i,E]$ of an interface $i$ and an embodiment $E$. Threads,
programs, services and so on are typical embodiments while the elements
of the aforementioned interface groups may act as interfaces. In the
financial setting embodiments are either true parts of an organization,
if no further formalization is performed or descriptions thereof which
specify their potential behavior. Such specifications can be cast as
processes and for instance be viewed as processes that may be specified
in detail in process algebra (see for instance the recent survey in
\cite{Fok00}) based formalisms like $\mu$CRL \cite{muCRL} or PSF
\cite{PSF}.

What it means for an entity $X$ that its behavior complies with a
financial transfer
interface $i$ is not easily defined with full precision, but informally it
is obvious: let $i$ be an interface in $\lifmB(E,A,M)$, then $X$
complies with $i$ if
\begin{itemize}
\item
all outgoing financial transfers of $X$ are instances of some (may
be more) positive interface elements $p$ that are contained in $X$
(i.e. $p \leq i$), and
\item
for all incoming transfer elements ${\sim} p$ of $i$ (${\sim} p
\leq i$) there is a range of incoming transfers for $X$ which cover all
reasonable instantiations of the atoms of $p$.
\end{itemize}

\subsection{Declared components and contained components}
Having this definition available a declared component is a pair $[i,X]$
of an interface and a financial behavior $X$ that complies with $X$. 

A closely related concept is that of a contained component. This is a
pair $[\overline{i},X]$ with $X$ a behavior such that (ii) above holds
w.r.t. $i$ and moreover: all outgoing  transfers which are not
instances of a positive $p$ contained in $i$ are forbidden (blocked,
disallowed) while also all incoming transfers that are not instances of
a negated interface element $-p$ of $i$ are forbidden. For components a
convincing definition of their interface exists: $I([i,X]) =
I([\overline{i},X]) = i$.

In the discussion and examples below contained components will not be
used and attention will be limited to declared components, which will
be called components because no confusion can arise. This is no real
restriction because for any constrained component $[\overline{i},X]$
the pair $[i,[\overline{i},X]]$ is a declared component which happens
to possess the same interface and the same behavior when restricted to
instantiations of interface elements and of negated interface elements
of $i$.  These descriptions are vague to the extent that the very
nature of instantiations of transfer interface elements is left open.

If $C$ is a declared component with interface $i$ and $i \leq j$ then
$[j,C]$ is a declared component as well.

\subsection{Financial transfer architectures}
\label{fta}
A named (declared) component is a pair $e{:}C$ with $C$ a component. A
named interface is a pair $e{:}i$ with $i$ an interface (taken in
$\ifmB(E,A,M)$ or in $\fifgB(E,A,M)$).

A sequence of named localized FT-interfaces
$e_1{:}i_1,...,e_n{:}i_n$
is a closed financial transfer architecture (CFTA) if
\[\phi_R(\sum_{1\leq k \leq n}\phi_{e_k}(i_k)) = 0.\]

The simplest example is this: $i_1= e_2.a_m$, $i_2 = {\sim} e_1.a_m$.
Then $\phi_R(\phi_{e_1}(e_2.a_m) + \phi_{e_2}({\sim} e_1.a_m)) =
e_2.a_m\at e_1 + ({\sim} e_1.a_m)\at e_2 = 0$. The reflection law has
been introduced precisely to make this kind of example work.

A realization of a CFTA consists of a sequence of named declared
components $e_1{:}[i_1,X_1],...,e_n{:}[i_n,X_n]$.

\subsection{A survey of components}
The behavior part of a component has been left unformalized in the
preceding definitions. There are many ways in which behaviors may be
conceived. For instance all transfers involved may be records of past
events. In that case the architecture describes an abstraction of a
bookkeeping. Alternatively a behavior may contain a tree of potential
unfoldings of future behavior (in other words a process in the sense of
process algebra or more generally in the sense of transition systems).
Yet another option is that both aspects are present in all
descriptions.

\section{Financial transfer interface design}
\label{sec:5}
Before providing examples the main expected merits of design and
specification, if not engineering, of financial transfer interfaces
(FTI's) may be listed.  Three types of artefacts may be engineered:
LFTI's (local FTI's), GFTI's (global FTI's, also called FTIA's for FTI
architectures, usually found by means of sums of globalized LTI's
following \ref{gigd}, and CFTIA's, for FTIA's that satisfy the 0-sum
criterion mentioned in \ref{fta}.

Precisely for formulating the 0-sum criterion the (commutative
additive) group structure of interfaces is considered helpful. An
alternative formulation is that this group structure provides multisets
with (multiple) negative occurrences as well as (multiple) positive
occurrences. Expected advantages of working with FTI's (including both
LFTI's and GFI's (=FTIA's)) include the following:

\begin{enumerate}
\item
An FTI provides qualitative information prior to any quantitative 
information. If an existing organization is analyzed FTI
design constitutes
a form of reverse engineering that ought to lead to an agreement.
Before such an agreement is achieved it may be pointless to proceed
with quantification
of financial streams. An FTI aggregates logical information about money 
streams to a comprehensible whole (in principle at least).
\item
Only once a CFTIA is known it is plausible and helpful to apply
Kirchhoff's current law for electrical  circuits \cite{Kirchhoff}
to the money streams that flow into and out of each entity
(the current entering any junction is equal to
the current leaving that junction).

\item
An FTI may be used for describing past transactions during a specified 
time interval, say a fiscal year, but it may also be used to provided
a qualitative 
perspective on expected expenditures and incomes.
But it may also be used for 
planning data, as planned transfer might be conceived
as a mode of transfer.
\item
FTI descriptions are independent of existing or expected financial
systems and theories.
\item
An FTI description is neutral concerning profit or loss because of the full
absence of quantitative information. But it provides an important
tool for setting the 
stage in advance if one is to analyze the effect of certain 
`profit centers' by providing
an incentive to be as clear as possible about the boundaries of
such entities.
\item
If a new organization is designed, or | what occurs
more frequently | an 
organization is changing its structure it may be helpful to
design an expected CFTIA for the organization.
\item
In particular if sourcing decisions (in-sourcing, out-sourcing, 
out-sourcing continuation, back-sourcing,
introduction of a shared service center) are contemplated a
precise analysis of the 
CFTIA before and after the implementation of the envisaged
sourcing decision
may be helpful.
This aspect relates FTI's to \cite{RD04} and \cite{D05}. 
\end{enumerate}

\subsection{What to expect from examples?}
This paper is not about a tool and what has been experienced by using
it, lessons from practice and so on. The story about FTI's has emerged
from working on a question (i), combined it with a short term objective
(ii) and a long term perspective (iii):

(i) How can anything 'logical' be said about finance? The motivation
for this question being that the notorious difficulty of designing a
clear language concerning financial matters may well be compared with
similar difficulties in computing.

(ii) An attempt to turn interface groups as proposed in \cite{BP06b}
into a useful tool for the investigation of IT outsourcing processes.

(iii) The working hypothesis that thread algebra (see \cite{BM04}) will
prove to be a significant concept for the specification of financial
systems, perhaps after an extension to a timed thread algebra. Extending
process algebra (see e.g. \cite{Fok00}) to timed versions has proven
feasible, see \cite{BaM02}, and the design of timed thread algebras is
definitely far simpler. The argument for this working hypothesis is
that the full complexity of arbitrary interleaving is not helpful in
initial stages of financial planning. Were financial planning to be
considered safety-critical in the way embedded computing is in
spacecraft then more general theories like process algebra and model
checking might come into play in full (and cumbersome) force.

Examples need to be given in this stage to demonstrate the reader that
working out an LFTI or an FTIA is both doable and potentially
informative. Demonstrating that striving for CFTIA's is of pragmatic
value can't be done by means of textual examples. That step follows
from the assumption that quantitative analysis needs a closed system
approach (at least at some level of abstraction) and that the 0-sum
criterion expresses that in an optimal way. But it may well be that the
main merit (if any) of designing LFTI's and (C)FTIA's lies in the
clarification that takes place during the design process of rather than
in obtaining a reliable stepping stone for moving towards a
quantitative model of an organization's financial processes.

\subsection{An example in detail}
The example draws from facts about our own academic institution, the
`Universiteit van Amsterdam'. The jargon has been provisionally
translated and the setting has been significantly simplified. Invisible
to readers but clearly recognizable for the authors is the circumstance
that deep differences of opinion can be spotted concerning the
appropriate LFTI's which are to be expected in a novel formal and
financial structure of the organization which is currently being
designed.

The default (own) institution (UvA) name is left implicit, within this
the default name (FS for faculty of science) is left implicit. There is
no need to work out the whole organization is equal detail for all of
its parts. This example is most specific in the aspects the authors
know best. Other people may add correspondingly precise descriptions of
their own parts of the organization, and a substantial task may then
remain if a CFTIA is to be manufactured from the set of these parts.

The example will focus on three entities:
 \texttt{HOSC06}, \texttt{MaEIis:SE}, and \texttt{MaEIis}.

\subsection{E, entities} 
The 'part of' relation between entities is not made explicit,
but transpires from the following naming scheme:

\begin{enumerate}
\item \texttt{FCsp}, facilities center: space
\item \texttt{FCeq}, facilities center: equipment
\item \texttt{FCrm}, facilities center: reproduction and media
\item \texttt{FCcat}, facilities center: catering
\item \texttt{FinC}, financial center
\item \texttt{ICs}, informatics center services
\item \texttt{ICc}, informatics center consultancy

\item \texttt{FS}, faculty of science, containing the following entities:
	\begin{itemize}
	\item \texttt{ESSC}, educational shared service center, containing
		\begin{itemize}
		\item \texttt{IO}, international office
		\item \texttt{SA}, student administration
		\item \texttt{CMD}, course material distribution
		\item \texttt{FM}, financial management
		\item \texttt{SC}, student counseling
		\item \texttt{TTP}, timetabling and planning
		\item \texttt{MC}, marketing and communications
		\end{itemize}
	\item \texttt{BaEIs}, bachelor Educational Institute (EI) of science
	\item \texttt{MaEIis}, master (Ma) EI of information sciences, 
	\item \texttt{MaEIes}, MaEI of exact sciences
	\item \texttt{MaEIles}, MaEI of life and earth sciences
	\item \texttt{MaEIps}, MaEI of professional studies
	\item \texttt{RIll}, Research Institute (RI) of logic and language
	\item \texttt{RIi}, RI of informatics, containing:
		\begin{itemize}
		\item \texttt{RIi:L:CSP} Lab (L) of computing, system
architecture and programming (CSP)
			\begin{itemize}
			\item \texttt{RIi:L:CSP:SE}, section software engineering (SE)
			\item \texttt{RIi:L:CSP:CSA}, section computer
systems architecture
			\item \texttt{RLi:L:CSP:SNE}, section systems and network engineering
			\item \texttt{RIi:L:CSP:CS}, section computational science
			\end{itemize}
		\item \texttt{RIi:L:HCS}, L of human-computer studies
		\item \texttt{RIi:L:IS}, L of intelligent systems
		\end{itemize}
	\item \texttt{RIapp}, RI for astroparticle physics
	\item \texttt{RIms}, RI for mathematics and statistics,
	\item \texttt{RIlsbe}, RI for life science: biodiversity and ecology
	\item \texttt{RIlsmb}, RI for life science: molecular biology
	\item \texttt{RIlsnh}, RI for life science: natural history
	\item \texttt{RIe}, RI for education in science
	\item \texttt{RIhep}, RI for high energy physics
	\item \texttt{RIep}, RI for experimental physics
	\item \texttt{RItp}, RI for theoretical physics
	\item \texttt{RIc}, RI for chemistry
	\item \texttt{Di}, division of informatics 
	\item \texttt{Dmap}, division of mathematics, astronomy and physics
	\item \texttt{Dc}, division of chemistry
	\item \texttt{Dles}, division of life and earth sciences
	\end{itemize}
\item \texttt{FH}, faculty of humanities
\item \texttt{FSBS}, faculty of social and behavioral sciences
\item \texttt{FL}, faculty of law
\item \texttt{FBE}, faculty of business and economy
\item \texttt{MS}, medical school
\item \texttt{MSd}, medical school for dentistry
\item \texttt{NWO}, national research funding organization 
\item \texttt{LSU}, local sister university 
\item \texttt{RSU1}, remote sister university 1
\item \texttt{RSU2}, remote sister university 2
\item \texttt{LUC1}, local university college (polytechnic) 1
\item \texttt{LUC2}, local university college 2
\item \texttt{OEEins}, other external educational institutions
\item \texttt{OERins}, other external research institutions
\item \texttt{OEo}, other external organizations
\item \texttt{OEind}, other external individuals (including staff)
\end{enumerate}

\subsection{A, modes} 
Only three modes are distinguished:
\begin{enumerate}
\item \texttt{cash}, cash payment
\item \texttt{it}, internal transfer
\item \texttt{et}, external transfer
\end{enumerate}

\subsection{M, motives} 
Motives capture both an abstraction of the service delivered and a qualification of the 
underlying service level agreement (SLA):
\begin{enumerate}
\item \texttt{hmt:csla}, hours multiplied by tariff (HMT) based on common service level agreement (SLA)
\item \texttt{hmt:nsla}, HMT based on negotiated SLA
\item \texttt{hmt:isla}, HMT based on incidental SLA
\item \texttt{hmt:rn}, HMT based on retrospective negotiation
\item \texttt{fp:dsla}, fixed price for dedicated SLA
\item \texttt{fp:fsla}, fixed price for flexible SLA
\item \texttt{fp:rn}, fixed price based on retrospective negotiation
\item \texttt{spe:rq}, staff personal expenditure compensation, retrospectively quantified
\item \texttt{spe:qa}, staff personal expenditure compensation, quantified in advance
\item \texttt{qmv:cp}, quantity multiplied by volume, common pricing
\item \texttt{qmv:ip}, quantity multiplied by volume, incidental pricing
\item \texttt{mbba}, model based budget allocation
\item \texttt{fbba}, (production) figures based budget allocation
\item \texttt{fbbr}, (production) figures based budget restitution
\item \texttt{us}, unspecified
\item \texttt{usr}, unspecified restitution
\end{enumerate}

\subsection{Examples of local interfaces}
Given these constants for the sorts $E$, $A$ and $M$, it is possible to
denote a vast number of local interfaces.
Beforehand it should
be stated that there is of course no unique mot plausible LFTI for any
entity.  The plausibility of a particular LFTI can only be judged in
the context of a coherent philosophy on how the organization as a whole
is supposed to function. But at the same time it can convey important
information about this philosophy.
As a first example consider \texttt{MaEIis}, while ignoring its partitioning. 

\texttt{LFTI4MaEIis0 =\\
RIll.it(hmt:csla + hmt:nsla + hmt:rn) + \\
RIi.it(hmt:csla + hmt:nsla + hmt:rn) +\\
FH.it(fp:nsla) + \\
FSB.it(fp:nsla) + \\
FBE.it(fp:nsla) + \\
FL.it(fp:nsla) +\\
ICc.it(fp:dsla) + \\
ESSC.it(fp:fsla) +\\
LSU.et(fp:dsla) + \\
LUC1.et(fp:dsla + fp:rn + us) +\\
OEo.et(fp:fsla) +\\
OEind.et(fp:dsla) +\\
2 x OEEins.et(fp:fsla + fp:dsla) +\\ 
FSs.it(qmv:cp + qmv:ip) +\\
FSrm.it(qmv:cp + qmv:ip) +\\
FScat.it(qmv:cp + qmv:ip) +\\
ICs.it(qmv:cp) +\\
ICc.it(hmt:csla) +\\
OEind.et(fp:dsla + spe:qa + spe:rq) \\
-FS.it(mbba + fbba + us) + FS.it(fbbr + usr)\\
-OEins.et(fp:dsla) + RIi.it(fp:dsla) + LSU.et(fp:dsla)
}

The second interface for \texttt{MaEIis} modifies the first one by
requiring services from sister faculties to be provided at hours times
tariff basis using a common SLA. Moreover it opens the possibility of
transfers to and from the division (though giving no clues as to the
motives for such transfers).  Moreover costs can be made for a
conference (NIOC07).  Cash payments can be received at the entrance
from those who did not register in advance and a cash payment may be
made to an invited speaker (whose credit card unexpectedly malfunctions
for instance).

\texttt{LFTI4MaEIis1 = LFTI4MaEIis0 \\
-FH.it(fp:nsla) + FH.it(hmt:csla)\\
-FSB.it(fp:nsla) + FSB.it(hmt:csla)\\
-FBE.it(fp:nsla) + FBE.it(hmt:csla)\\
-FP.it(fp:nsla) +  FH.it(hmt:csla) +\\
-Di.it(us) + Di.it(usr) +\\
-OEind.cash(us) + NIOC07.et(us) + OEind.cash(us)}

The picture may be further refined by splitting the \texttt{MaEIis}
into the various components that have been listed. But that might be
considered artificial, as it will not introduce any new types of
transfers. 
On the other hand,
 if it is considered preferable to allocate all incoming
funds to one of the master programs or to G (management and planning),
this can be done for instance just for one program (say \texttt{SE})
and one obtains e.g.

\texttt{LFTI4MaEIis2 = LFTI4MaEIis1 +\\
SE.it(mbba + fbba + us) \\
- OEEins.et(fp:fsla + fp:dsla) - ICc.it(fp:dsla)}

The transfers specific for running the program \texttt{SE} will now
feature as a part of the LFTI for SE. Details of that LFTI will not be
presented as an example, assuming that the reader can imagine how that
might work. It should be stressed that these interface descriptions are
quite realistic but still require significant additional explanation.
As a consequence it can be noted that extensive comments are essential
if LFTI specifications are intended to be practically helpful in any
concrete case. One might include comments in a LaTeX like environment
description: \texttt{\%[.....\%]} which should follow directly the
interface element that is being commented. In the example below
\texttt{LFTI4MaEIis0} is split in a part with comments and a part
without comments.

\texttt{LFTI4MaEIis0 = LFTI4MaEIis0comm + LFTI4MaEIis0nocomm\\
~\\
LFTI4MaEIis0comm =\\
RIll.it(hmt:csla + hmt:nsla + hmt:rn) +\\
\mbox{~}\%[full cost compensation (FCC) for RIll teaching staff (TS)\%] \\
RIi.it(hmt:csla + hmt:nsla + hmt:rn) + \%[FCC for RIi TS\%]\\
FH.it(fp:nsla) ~+ \%[negotiated compensation (NC) for FH TS\%]\\
FSB.it(fp:nsla) + \%[NC for FSB TS\%]\\
FBE.it(fp:nsla) + \%[NC for FBE TS\%]\\
FL.it(fp:nsla) ~+ \%[NC for FL TS\%]\\
ICc.it(fp:dsla) + \%[NC for ICc TS\%]\\
ESSC.it(fp:fsla)+ \%[fixed price for flexible realization of SLA\\
LSU.et(fp:dsla) + \%NC for FH TS\%]\\
LUC1.et(fp:dsla + fp:rn + us) + \%[NC for FH TS \& \\
\mbox{~} compensation for services without preceding SLA \& \\
\mbox{~} allocation for joint management effort  on ASICT\&\\
\mbox{~} compensation for use of SNE laboratory space\%]\\
~\\
LFTI4MaEIis0nocomm =\\
OEo.et(fp:fsla) +\\
OEind.et(fp:dsla) +\\
2 x OEEins.et(fp:fsla + fp:dsla) +\\ 
FSs.it(qmv:cp + qmv:ip) +\\
FSrm.it(qmv:cp + qmv:ip) +\\
FScat.it(qmv:cp + qmv:ip) +\\
ICs.it(qmv:cp) +\\
ICc.it(hmt:csla) +\\
OEind.et(fp:dsla + spe:qa + spe:rq) \\
-FS.it(mbba + fbba + us) + FS.it(fbbr + usr)\\
-OEins.et(fp:dsla) + RIi.it(fp:dsla) + LSU.et(fp:dsla)
}

\section{Discussion and concluding remarks}
\label{sec:6}
After a discussion of related literature concerning interfaces and
a discussion of related financial literature some directions for
subsequent work are mentioned.

\subsection{Interfaces and interface groups in other work}
The term interface group has been discussed in \cite{OOW91} and occurs
widely in the literature about internet protocols; it was used by 
Keith Cheverst et. al. in the context of groupware description
\cite{CBDF99}.
These uses of the phrase make no reference to the mathematical theory
of groups. For that reason we consider it justified to propose the
meaning assigned to `interface group' in this paper for use in a
theoretical context.

A significant theory of interfaces and components is given by Scheben
in \cite{Sch05}.  Issued requests are referred to as  `required
services', whereas accepted requests are referred to as provided
services. Scheben also designs a general notation for the description
of component interfaces.  In \cite{VMDOL05} interfaces are cast in
terms of interface automata.  What is called a reply service in
\cite{BM05} is a special case of interface automata.

A convincing example of interfaces is given by the so-called
instruction set architectures for microprocessors, used throughout
computer engineering, which can be given a theoretical basis by means
of the classical theory of Maurer in \cite{M66}. Recent work on
improved architectures depends on generic transformations of
instruction sets (see \cite{JL00}).

The constraint that a boolean is given in return when a request $a$ is
accepted ($-a/TF$) may be considered a `promise'.  Mark Burgess has
been developing  theory of promises for the description of services in
networks of autonomous components (see \cite{Bu05}).  Interface groups
can be used to formalize parts of his work. Besides in general systems
architecture for computing and in the foundations of bookkeeping, as
suggested in this paper, interface groups might be used to formalize
the work on sourcing architectures by Rijsenbrij and Delen in
\cite{RD04} and subsequently in \cite{D05}.  Their theory of atomic
outsourceable units requires a formal foundation which critically
depends on a systematic use of interfaces.

\subsection{Further questions concerning financial transfer architectures}
Many further projects can be imagined, in particular with modeling
increasingly more complex organizations by means of specifications of FT
interfaces that describe their internal architecture. Unavoidably for
complex organizations these interfaces will not be static but may
change in time. The specification of dynamically changing interfaces
poses some challenge and can't simply be imported from the computer
science literature.

We have until now failed to find related literature in management finance
theory if any exists. Finding appropriate connections with
theories of finance is an objective that will require further attention in
subsequent research.


\begin{thebibliography}{99}

\bibitem{BaM02}
J.C.M Baeten and C.A. Middelburg,
\newblock \emph{Process algebra with timing}.
\newblock Springer-Verlag, 2002

\bibitem{BB03}
J.A.~Bergstra and I.~Bethke.
\newblock Polarized process algebra and program equivalence.
\newblock In J.C.M.~Baeten, J.K.~Lenstra, J.~Parrow, and
G.J.~Woeginger (editors), {\em Automata, Languages and Programming, 30th
International
Colloquium, ICALP 2003, Eindhoven, The Netherlands, June 30 - July 4},
Springer-Verlag, LNCS 2719:1-21, 2003.

\bibitem{BHK90}
J.A. Bergstra, J. Heering and P. Klint.
\newblock Module Algebra.
\newblock {\em Journal of the ACM}, 37(2):335-372, 1990.

\bibitem{BL02}
J.A. Bergstra and M.E. Loots.
\newblock Program algebra for sequential code.
\newblock {\em Journal of Logic and Algebraic Programming},
 51(2):125-156, 2002.

\bibitem{BM04}
J.A. Bergstra and C.A. Middelburg,
\newblock Thread algebra for strategic interleaving.
\newblock Technical report PRG0404, Programming Research Group,
University of Amsterdam, November 2004. 
\newblock To appear in \emph{Formal Aspects of Computing}.


\bibitem{BM05}
J.A. Bergstra and C.A. Middelburg. Thread algebra with
multi-level strategic interleaving.
\emph{Theory of Computing Systems}, 41(1):3-32, 2007.

\bibitem{BP06}
J.A. Bergstra and A. Ponse.
\newblock Execution architectures for program algebra.
\newblock {\em Journal of Applied Logic}, 5(1):170-192, 2007.

\bibitem{BP06b}
J.A. Bergstra and A. Ponse.
\newblock Interface Groups for Analytic Execution Architectures,
\newblock {\em PRG Electronic Report PRG0601}, 
Programming Research Group, 
Department of Computer Science, 
University of Amsterdam, 
 2006.

\bibitem{Bu05}
M. Burgess.
\newblock An approach to understanding policy based on autonomy 
and voluntary cooperation.
\newblock {\em 16th IFIP/IEEE Distributed Systems Operations and 
Management (DSOM 2005)}, Springer-Verlag, LNCS 3775, 2005.

\bibitem{CBDF99}
K. Cheverst, G. Blair, N. Davies and A. Friday.
\newblock The support of mobile-awareness in collaborative groupware.
\newblock {\em Personal Technologies}, 3(1-2):33-42, 1999.

\bibitem{D05}
G.P.A.J. Delen.
\newblock (In Dutch)
{\em Decision- en controlfactoren voor sourcing van IT.}
\newblock Ph. D. Thesis, University of Amsterdam
(van Haren publishing Zaltbommel), 2005.

\bibitem{Fok00}
W.J. Fokkink.
\newblock {\em Introduction to Process Algebra}.
\newblock Texts in Theoretical Computer Science.
Springer-Verlag, 2000.

\bibitem{muCRL}
J.F. Groote and A. Ponse.   The syntax and semantics of muCRL.
In A. Ponse, C. Verhoef, and S.F.M. van Vlijmen (editors), 
\emph{Algebra of Communicating Processes}, Utrecht 1994. Workshops in
Computing, Springer-Verlag, pages 26-62, 1995.
(See also \url{http://homepages.cwi.nl/~mcrl/}).

\bibitem{JL00}
C.R.~Jesshope and B.~Luo.
\newblock Micro-threading, a new approach to future RISC.
\newblock {\em In ADAC 2000, IEEE Computer Society Press},
34-41, 2000.

\bibitem{M66}
W.D. Maurer.
\newblock A theory of computer instructions.
\newblock {\em Journal of the ACM}, 13(2):226-235, 1966.
An extended version appeared in
\emph{Science of Computer Programming}, 60(3):244-273, 2006.

\bibitem{PSF}
S. Mauw, G.J. Veltink. A Process Specification Formalism, 
\emph{Fundamenta Informaticae}, XIII:85-139, 1990.
(See also \url{http://www.science.uva.nl/~psf/}).

\bibitem{OOW91}
M. Olsen, E. Oskiewics and J. Warne.
\newblock A model for interface groups.
\newblock {\em JProc. 10th IEEE Symp. on Reliable Distributed Systems},
98-107, (Pisa Italy) 1991.

\bibitem{PZ}
A. Ponse and M.B. van der Zwaag.
\newblock An introduction to program and thread algebra. In
A. Beckmann et al. (editors), \emph{CiE 2006}, LNCS 3988, pages 445-458,
Springer-Verlag, 2006.

\bibitem{RD04}
D.B.B. Rijsenbrij and G.P.A.J. Delen.
\newblock (In Dutch)
Enterprise-architectuur is een noodzakelijke voorwaarde 
voor verantwoorde outsourcing.
\newblock {\em In:  IT service management best practices}
(red. J. van Bon), 
van Haren Publishing Zaltbommel, 35-58, 2004

\bibitem{Sch05}
U.~Scheben.
\newblock Hierarchical composition of industrial components.
\newblock {\em Science of Computer Programming}, 56:117-139, 2005.


\bibitem{Kirchhoff}
C.R. Paul.
\newblock \emph{Fundamentals of Electric Circuit Analysis}.
John Wiley \& Sons, 2000.

\bibitem{VMDOL05}
P. V\"olgyesi, M. Mar\'oti, S. D\'ora, E. Osses, and \'A.
L\'edeczi.
\newblock Software composition and verification for sensor networks.
\newblock {\em Science of Computer Programming}, 56:191-210, 2005.

\end{thebibliography}
\end{document}